%% file: Main.tex
\documentclass[twocolumn]{aastex7} %
\hypersetup{linkcolor=red,citecolor=blue,filecolor=cyan,urlcolor=magenta}
\usepackage{longtable}
\usepackage{array}
\usepackage{booktabs}
\usepackage{caption}
\usepackage{siunitx}
\usepackage{amsmath}
\usepackage{graphicx}
\usepackage{caption}
\usepackage{subcaption}
\usepackage{comment}
\newcommand{\Teff}{$T_{\mathrm{eff}}$} 

\newcommand{\kms}{km$\,\rm s^{-1}$}

\begin{document}

\title{Lithium in Wide Binaries: Effective Temperature Governs Depletion while Rotation Plays a Minor Role}

\author{Cheng-Cheng, Xie}
\affiliation{School of Sciences, HangZhou Dianzi University, HangZhou 310018, People's Republic of China}
\affiliation{Zhejiang Branch of National Astronomical Data Center, Hangzhou 310018, China} 
\email{hjtian@hdu.edu.cn}
\author{Hai-Jun, Tian}
\affiliation{School of Sciences, HangZhou Dianzi University, HangZhou 310018, People's Republic of China}
\affiliation{Zhejiang Branch of National Astronomical Data Center, Hangzhou 310018, China} 
\email{hjtian@hdu.edu.cn}
\author{Jian-Rong Shi}
\affiliation{National Astronomical Observatories, Chinese Academy of Sciences, Beijing 100101, }
\affiliation{University of Chinese Academy of Sciences, Beijing 100049, People's Republic of China}
\email{sjr@bao.ac.cn}
\author{Ze-Ming Zhou}
\affiliation{National Astronomical Observatories, Chinese Academy of Sciences, Beijing 100101, }
\affiliation{University of Chinese Academy of Sciences, Beijing 100049, People's Republic of China}
\email{sjr@bao.ac.cn}

\correspondingauthor{Hai-Jun Tian, Jian-Rong Shi}
\email{hjtian@hdu.edu.cn, sjr@bao.ac.cn}




\begin{abstract}
Using a sample of 116 wide binary systems as coeval and chemically homogeneous stellar pairs, we investigate the factors governing lithium depletion in main-sequence stars. We recover the well-established morphology of the lithium--effective temperature (\Teff) relation, including the Li dip (6200--6600\,K), the Li plateau (6000--6200\,K), and a linear trend for cooler stars (\Teff $<$ 6000\,K), where lithium abundance increases by $\sim$0.15\,dex per 100\,K. We demonstrate that the apparent correlation between projected rotational velocity ($v\sin i$) and lithium abundance is secondary to the underlying \Teff\ dependence; $v\sin i$ is not an independent driver of lithium depletion in our sample. Notably, we identify an anomalous system within the Li dip where the primary star exhibits a $\sim$1.4\,dex lithium excess compared to its secondary companion at nearly identical \Teff. We discuss two plausible origins for this anomaly: external enrichment via planetesimal accretion or planetary engulfment, and binary interactions with an unresolved tertiary companion. Our results confirm \Teff\ as the dominant parameter controlling lithium depletion, while highlighting that additional, non-standard processes can occasionally produce significant lithium enrichment.

\end{abstract}

\keywords{Wide Binaries (251) --- Lithium Abundance --- History of astronomy(1868) --- Interdisciplinary astronomy(804)}

\section{Introduction} \label{sec:intro}

Lithium (Li) is a key diagnostic element in stellar astrophysics. Its surface abundance, denoted as \textit{A}(Li)\footnote{$\textit{A}(\mathrm{Li}) = \log [N(\mathrm{Li}) / N(\mathrm{H})] + 12$}, is readily destroyed by proton capture at temperatures exceeding $2.5 \times 10^{6}$\,K via $^7$Li(p,$\alpha$)$^4$He \citep{Burbidge1957, Bodenheimer1965}. Consequently, Li survives only in the outer layers of stars, making it a sensitive tracer of internal mixing and transport processes that connect the surface to hotter interiors. Standard stellar evolutionary theory \citep[SSET;][]{Deliyannis1990}, which primarily considers convection and pre-main-sequence burning, for G-type dwarfs such as the Sun and cooler K-type dwarfs, models predict significant early lithium depletion, as their deep convective zones transport surface material into hot interior regions where lithium is rapidly destroyed via nuclear burning. This process is expected to occur predominantly during the pre-main sequence phase. In contrast, hotter F- and A-type stars possess substantially shallower convective envelopes; consequently, SSET predicts that their surface lithium abundances should remain relatively unchanged throughout their main-sequence lifetimes. However, SSET fails to reproduce key observational patterns of Li depletion in main-sequence stars. These include the pronounced ``Li Dip'' in mid-F dwarfs \citep{Boesgaard1986,Balachandran1995}, the relatively constant ``Li Plateau'' in late-F to early-G dwarfs, and the severe mass-dependent depletion observed in cooler G- and K-dwarfs \citep{Cummings2017}. Observations of Li in open clusters of different ages show that G dwarfs continue to deplete Li during the main sequence, in sharp contrast to the SSET \citep{Soderblom1993,Pinsonneault1997}. The discrepancy between the SSET predictions and observations reveals the action of additional physical mechanisms not included in standard models.

To explain these discrepancies, rotational mixing induced by stellar spin-down has emerged as a leading candidate for the dominant non-standard Li depletion mechanism across spectral types \citep{Pinsonneault1989, Ryan1995, Pinsonneault1997, Sun2023}. The loss of angular momentum drives internal circulation and shear instabilities, which transport surface Li to deeper, hotter regions, where it is destroyed. This framework is supported by correlations between Li abundance and rotation, where slower rotators typically show higher Li depletion \citep{Pinsonneault1997}. Other processes such as atomic diffusion \citep{Richer1993}, mixing induced by internal gravity waves \citep{GarciaLopez1991ApJ}, and matter accretion \citep{LiXF2025ApJ} may also contribute to the observed scatter in Li abundances at a given effective temperature (\Teff) and metallicity. Additional evidence comes from the correlation of spindown to Li depletion \citep{Steinhauer2025}, and from stars such as short period tidally locked binaries with higher Li than normal, single stars \citep{Ryan1995}. Inside stars, Li, Be, and B survive to progressively deeper layers. By far the most powerful observational constraints on physical processes affecting Li depletion come from incorporating information from Be and B abundances as well, especially from the observed ratios of these elements \citep{Boesgaard2020, Deliyannis1998, Boesgaard2005}.

The efficiency of all these depletion mechanisms is fundamentally modulated by the structure of the stellar interior, particularly the depth of the surface convection zone \citep[SCZs,][]{Pinsonneault1997}. The depth of SCZ increases with decreasing stellar mass (and \Teff), leading to more efficient transport of Li to its destruction region. This explains the general trend of increasing Li depletion from F- to K-dwarfs. The Li Dip (approximately 6200--6800\,K) corresponds to a narrow mass range where the SCZ reaches a critical depth, often coupled with the onset of efficient rotational braking \citep{Theado2003}, resulting in severe Li depletion. The Li Plateau (approximately 6000--6200\,K) represents a transition regime where the SCZ is relatively shallow, allowing Li to be preserved at near-primordial levels in some populations. In cooler stars (\Teff $<$ 6000\,K), the deep SCZ leads to efficient Li destruction throughout the main sequence.

Metallicity further influences Li depletion through multiple channels. Higher metallicity increases atmospheric opacity, leading to a deeper SCZ at a given \Teff, which enhances convective mixing and the destruction of Li. Additionally, metallicity affects stellar structure, rotational evolution, and the efficiency of atomic diffusion processes \citep{Pinsonneault1997, Ryan2000}. These metallicity-dependent effects contribute to the observed variations in Li patterns between metal-rich and metal-poor stellar populations.

Wide binaries (WBs), defined as gravitationally bound stellar pairs with large separations ($\gtrsim$100\,AU), provide a powerful laboratory for isolating the effects of these physical mechanisms. Components of a WB are coeval and share a common initial chemical composition, having formed from the same molecular cloud core. High-resolution spectroscopic studies have robustly confirmed their chemical homogeneity, particularly for refractory elements \citep{Hawkins2019}. However, despite these identical initial conditions, WB components often exhibit significant differences in their surface Li abundances. These differences must therefore arise from internal stellar properties that have diverged during their evolution, such as mass (and hence \Teff), rotation, and convective efficiency—precisely the parameters that govern Li depletion. This makes WBs an ideal testbed for disentangling the various proposed depletion mechanisms, free from the confounding variables of age and initial composition that affect comparisons between field stars or even star clusters.

Recent studies of large samples of co-moving stellar pairs have quantified the intrinsic scatter in \textit{A}(Li) at fixed \Teff\ and metallicity, finding values of $\sim$0.35\,dex for solar-type stars \citep{Sun2024}. This scatter likely reflects the complex interplay of rotation, magnetic activity, and other internal processes. In particular, such studies also suggest that potential Li enrichment signatures from rare events such as planetary engulfment are often of similar amplitude to this intrinsic scatter, making them challenging to distinguish observationally \citep{Sevilla2022, Sun2024}. Therefore, analyzing systems with extreme Li anomalies, such as a component within a coeval pair showing an unusually high abundance, becomes crucial for identifying clear signatures of non-standard processes.

In this work, we perform a detailed analysis of lithium abundances in a sample of WBs. Our primary goals are to: (1) map the \textit{A}(Li)-\Teff\ relation in a coeval chemically homogeneous environment; (2) investigate the roles of \Teff\ (mass) and rotation in driving depletion; and (3) identify and scrutinize any systems exhibiting anomalous Li abundances that may point to specific events such as external enrichment. Our sample is constructed from \textit{Gaia} EDR3 wide binary catalogs, with atmospheric parameters and high-resolution spectra drawn from GALAH DR3 \citep{GALAHDR3}.

This paper is structured as follows. Section~\ref{sec:style} describes the selection of our WB sample and the spectroscopic data. Section~\ref{sec:abundances} details the methodology for determining Li abundances via spectral synthesis. Our main results, including the \textit{A}(Li)-\Teff\ pattern and the analysis of an anomalous system, are presented in Section~\ref{sec:results}. In Section~\ref{sec:DISCUSSION}, we discuss the implications of our findings and propose mechanisms to explain the observed Li anomaly. Finally, our conclusions are summarized in Section~\ref{sec:CONCLUSION}.

\begin{figure*}[!t]
\includegraphics[width=0.48\textwidth, trim=0cm 0.cm 0cm 0cm, clip]{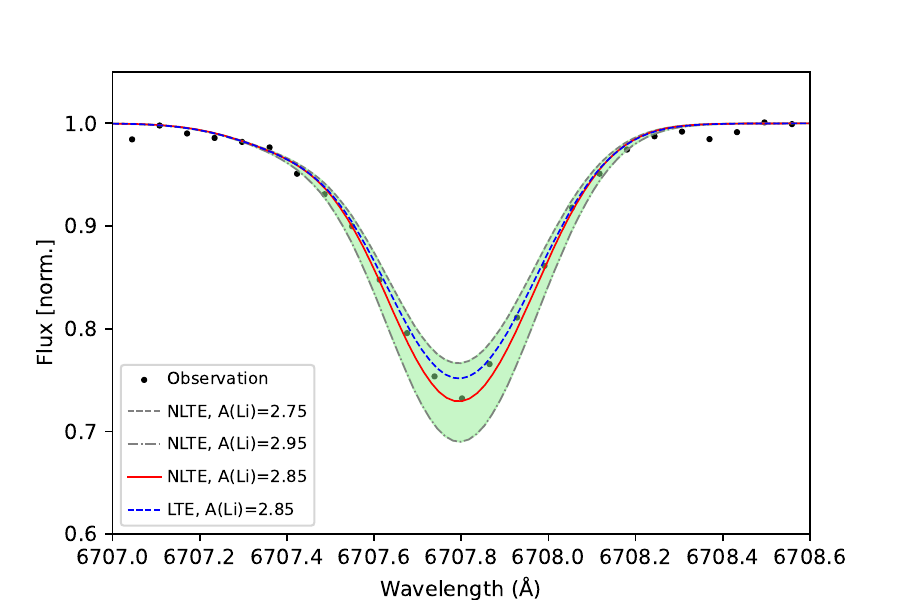}
\includegraphics[width=0.48\textwidth, trim=0cm 0.cm 0cm 0cm, clip]{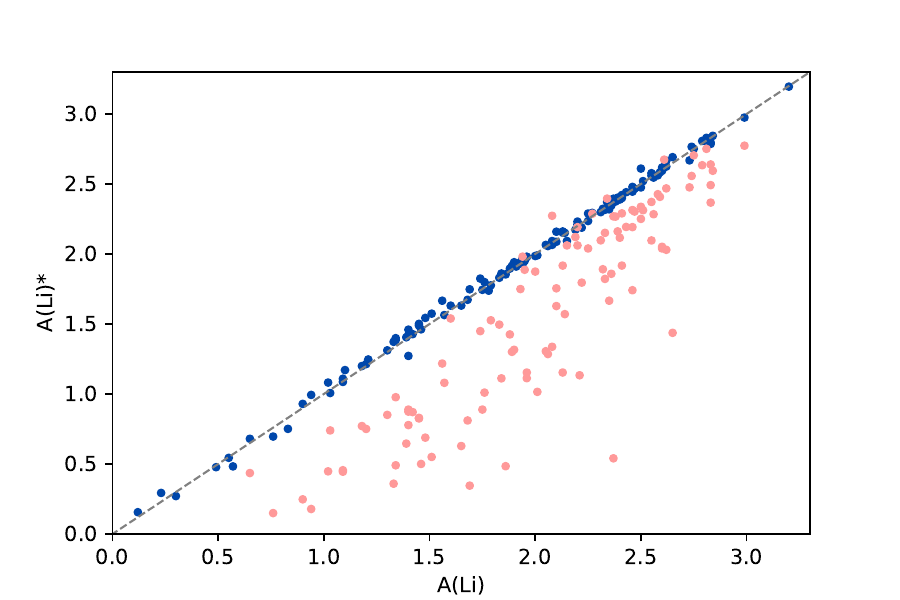}
\caption{Left panel: An example for {\it A}(Li) measurement. The black points are the observed spectra, and the red line is the best fit synthetic spectra with {\it A}(Li)=2.85\,dex. The gray dashed and dotted lines are the uncertainties of the fitting with {\it A}(Li)=2.75\,dex and 2.95\,dex, respectively. Right panel: The abscissa represents the Li abundance determined by us, while the ordinate represents the Li abundance provided by GALAH DR3 (blue points) or GALAH DR4 (red points) .}\label{fig:placeholder}
\end{figure*}

\section{DATA and Analysis}\label{sec:style}
\subsection{The Initial Wide Binaries Sample from \textit{Gaia}} 

This study uses the initial catalog of wide binary candidates identified by \citet{El-Badry2021} from \textit{Gaia} Early Data Release 3 \citep[\textit{Gaia} EDR3,][]{GaiaCollaboration2021}. The selection of candidates was constrained to sources nominally within a distance limit of 1\,kpc. The identification methodology closely followed the approaches established by \citet{El-Badry2018} and \citet{Tian2020}. This involved applying stringent criteria based on the parallax, proper motion, astrometric solution quality, and photometric properties of both stellar components within each candidate pair. Subsequent processing steps aimed to mitigate contamination by excluding cluster members, background projection pairs, and higher-order multiples (e.g., triples). For details on specific selection thresholds and filtering procedures, refer to Section 2 of \citet{El-Badry2021}. We use the 877,416 high bound probability Main Sequence - Main Sequence (MSMS) wide binaries as our initial sample.


\subsection{Wide Binaries Cross-matched with GALAH}

The Galactic Archeology with HERMES (GALAH) survey’s Data Release 3 \citep[DR3;][]{GALAHDR3} provides one-dimensional spectra, stellar atmospheric parameters, and individual elemental abundances for 678,423 high resolution spectra (R $\approx$ 28000) of 588,571 individual objects. They were observed with the HERMES spectrograph at the Anglo-Australian Telescope. The survey provides uniformly derived stellar parameters, such as effective temperature (\Teff), surface gravity (log $g$), metallicity ([Fe/H]), micro-turbulent velocity ($\xi_{t}$), $v\sin i$, and $R_V$, they were derived by using a modified version of the spectrum synthesis code Spectroscopy Made Easy (SME) and 1D MARCS model atmospheres. We got 218 pair spectra by cross-matching the initial MSMS binary sample with the GALAH DR3 catalogue. In addition, the GALAH DR3 provided the strict data quality flags (the values of those flags equal to 0 means that the parameters are reliable) of the stellar parameters and elemental abundances, to ensure the accuracy of the results. In this paper only the spectra quality parameters of "flag\_sp=0" and "fe\_h\_sp=0" have been considered.

We have limited the difference in $\Delta R_V$ (and $\Delta$[Fe/H]) (i.e., the difference $R_V$ or [Fe/H] between the two components of each binary), and removed the sources of $\Delta R_V$ (and $\Delta$[Fe/H]) beyond $\pm$3 $\sigma_{R_V}$ (and $\pm$3 $\sigma_{\rm[Fe/H]}$). Finally, we obtained a sample of 116 binary systems, and their observational properties and atmospheric parameters are listed in Table~\ref{tab:table1}.


Although GALAH DR3 and DR4 provided valuable  homogeneously derived catalogs of stellar parameters and elemental abundances—including Li \citep{GALAHDR3,buder2024galah}, we would like to perform an independent measurement of Li abundance for this study. This choice is prompted by noted inconsistencies between the two data releases, particularly a reported decrease in precision for certain abundances in DR4 \citep{buder_2025galahsurveydatarelease}. The differences stem from a substantial change in methodology: GALAH DR3 employed classical spectral synthesis with Spectroscopy Made Easy \citep{piskunov2017, buder2021}, while DR4 adopted a neural-network approach trained on synthetic spectra \citep{kane2025}. In both data releases, there exist some stars for which effective Li abundances are not provided. Due to these reasons, the Li abundances presented in the remainder of this work are based on our own dedicated spectral re-analysis.


\section{Measurement of Lithium Abundance}
\label{sec:abundances}

\subsection{Li abundance}
We adopt the stellar atmospheric parameters of stars, i.e., the effective temperature \Teff, surface gravity $\log g$, metallicity [Fe/H], and microscopic turbulence velocity $\xi_t$, from GALAH DR3 \citep{GALAHDR3}.
The Li abundances are derived under both the local thermodynamic equilibrium (LTE) and non-local thermodynamic equilibrium (NLTE) assumptions using the spectral synthesis method.  The LTE and NLTE Li abundances are determined from the resonance line at 6708\,{\AA} via the spectral synthesis method, which interactively finds the best matches between observed and synthetic spectra. Synthetic line profiles are calculated using the IDL/Fortran-based Spectrum Investigation Utility \citep[SIU,][]{1999PhDT.......216R} program with the 1D LTE MARCS atmospheric models \citep{2008A&A...486..951G}. In our analysis, the adopted Li line data and the atomic model have been presented in detail by \cite{2007A&A...465..587S}. To solve the coupled radiative transfer and statistical equilibrium equations, a revised DETAIL code \citep{Mashonkina2011} is employed, which uses an accelerated $\Lambda$-iteration method \citep{1991A&A...245..171R,1992A&A...262..209R}. During the fitting procedures, free parameters are the Li abundance and a Gaussian profile, which accounts for instrumental, rotational, and macro-turbulent broadening.
Figure~\ref{fig:placeholder} (left panel) shows an example of the spectral synthesis fitting. 
The measured \textit{A}(Li) of the 116 binary systems is given in Table~\ref{tab:table1}, including 61 (65) systems of which \textit{A}(Li) is not provided by GALAH DR3 (DR4). 

\subsection{The Uncertainties of Our Li Abundances}

The abundance of Li is sensitive to stellar atmospheric parameters, in particular of temperature. Thus, it is important to investigate the uncertainty in Li abundance as a result of the uncertainties in the stellar parameters. The uncertainty of Li abundance is quantified by the changes in their errors for the four stellar parameters, i.e., \Teff, log $g$, [Fe/H], and $\xi_{\rm t}$. As an example, the uncertainties of the abundance of Li due to the errors in the stellar parameters for WB05a are shown in Table~\ref{tab:ali_uncertainty}. The results indicate that, among the considered sources of uncertainty, errors in \Teff\ have the most significant impact on $A(\mathrm{Li})$, while the contributions from other parameters are negligible within their respective uncertainty ranges. For our sample, where the typical \Teff\ uncertainty is 70--80\,K, this propagates to a measurement uncertainty in $A(\mathrm{Li})$ of approximately 0.05--0.06\,dex.  


\begin{table}[h]
  \centering
  \caption{The uncertainties of measured {\it A}(Li) due to the errors in the stellar parameters of WB05a}
  \label{tab:ali_uncertainty}
  \small 
  \setlength{\tabcolsep}{2.5pt} 
  \begin{tabular}{c *{8}{c}}
    \toprule
    Atmo. & 
    \multicolumn{2}{c}{\Teff} & 
    \multicolumn{2}{c}{log $g$ } & 
    \multicolumn{2}{c}{[Fe/H]} & 
    \multicolumn{2}{c}{$\xi_{\rm t}$} \\
     & \multicolumn{2}{c}{(K)} & \multicolumn{2}{c}{(dex)} & \multicolumn{2}{c}{(dex)} & \multicolumn{2}{c}{ (\kms)} \\
    \midrule
    Mean & \multicolumn{2}{c}{5892} & \multicolumn{2}{c}{4.2} & \multicolumn{2}{c}{0.3} & \multicolumn{2}{c}{1.2} \\
    Error & $+$72 & $-$72 & $+$0.2 & $-$0.2 & $+$0.05 & $-$0.05 & $+$0.1 & $-$0.1 \\
    $\Delta A({\rm Li})$ & $+$0.06 & $-$0.05 & $+$0.01 & 0.00 & $+$0.01 & 0.00 & 0.00 & 0.00 \\
    \bottomrule
  \end{tabular}
  \normalsize 
  \tablecomments{The uncertainties in $A(\mathrm{Li})$ are derived from propagating the errors in stellar parameters through spectral synthesis analysis.}
\end{table}

\subsection{The Validation of Our Li Abundances}

To validate the measured \textit{A}(Li), we compare the derived \textit{A}(Li) with those from GALAH DR3 and DR4. The right panel of Figure \ref{fig:placeholder} shows that our \textit{A}(Li) is well in agreement with the values of GALAH DR3, while systematically higher than those of GALAH DR4. This comparison confirms that our measured \textit{A}(Li) is reliable.

\section{Results}
\label{sec:results}

The coeval and co-natal nature of WBs ensure that the component stars share consistent metallicity and age. As the fundamental parameters are well-constrained, WBs offer an ideal laboratory for investigating which factors will influence the stellar Li abundance {\it A}(Li). 

\begin{figure*}[!t]
\centering
\includegraphics[width=1.0\textwidth, trim=2.0cm 0.cm 3.5cm 1cm, clip]{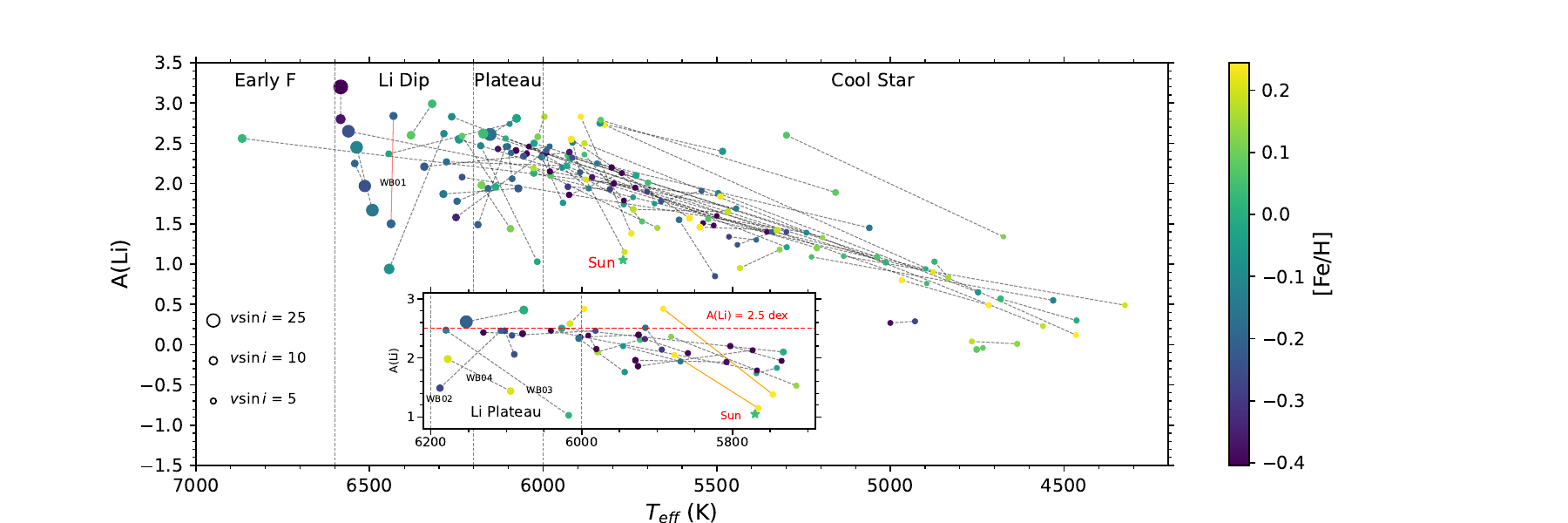}
\includegraphics[width=1.0\textwidth, trim=2.0cm 0.cm 3.5cm 1.05cm, clip]{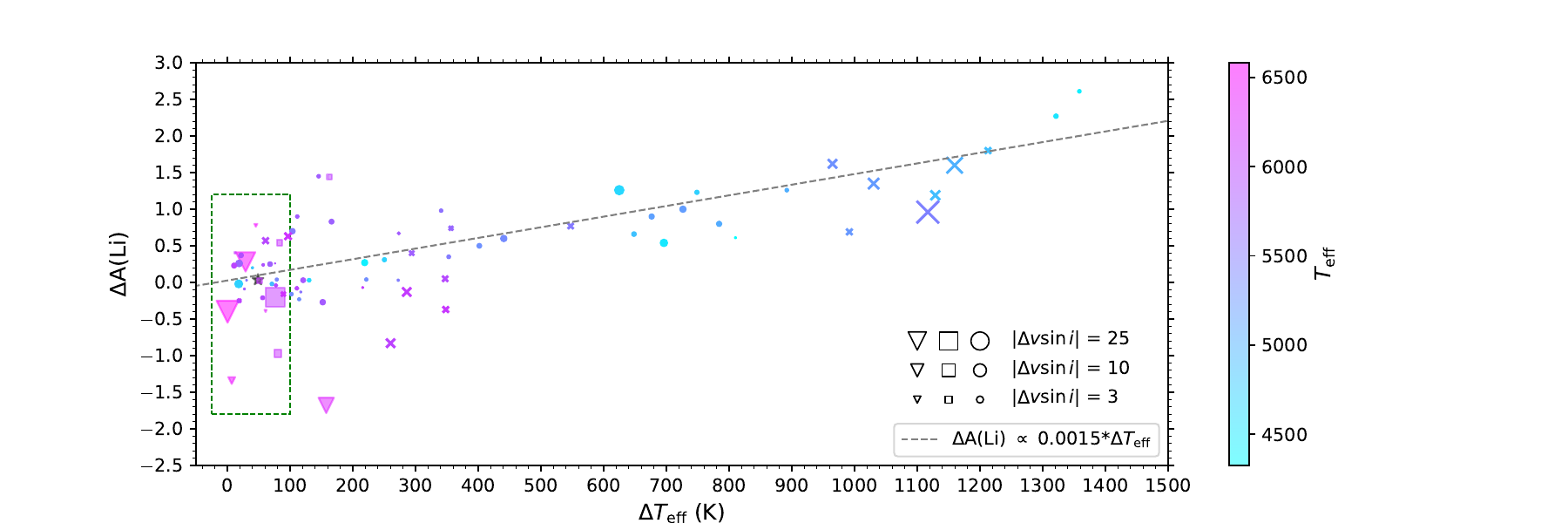}
\caption{Upper: The 1D NLTE {\it A}(Li) – $T_{\rm eff}$ pattern for 116 pair WBs. Each binary pair is connected by a black dashed lines. Component stars are marked by filled circles which scaled in size according to their $v\sin i$ values, and color-coded by their [Fe/H], spanning the range of $\langle\rm[Fe/H]\rangle \pm 1.5\sigma$ (approximately $-0.18 \pm 0.32$\,dex). The inset zooms in on binary pairs where both components have \Teff\ between 5700\,K and 6200\,K. Bottom: $\Delta$A(Li) as a function of $\Delta$\Teff\ between the primary and the secondary of each pair binary. The symbols are colored by the \Teff\ of the secondary stars. The symbols size are coded by the $|\Delta$$v\sin i|$ between the primary and secondary stars. The dashed black line represents a linear fit to the filled points (cool stars only), excluding symbols for Li dip binaries (inverted triangles), Li plateau binaries (filled squares), and systems where only the secondary is a cool star (crosses). The green dished box includes the binary pairs with negligible \Teff\ differences ($\Delta$\Teff$<100$\,K). The black asterisk marks the median value of A(Li) in the box.}\label{fig:offset2}
\end{figure*}

Figure~\ref{fig:offset2} plots Li abundances as a function of the effective temperature for our binary sample. The upper panel presents {\it A}(Li) against \Teff\ for all 116 systems, with each binary pair connected by a black dashed line. Individual components are represented by filled circles, color-coded by [Fe/H] and scaled in size according to their $v\sin i$ values. The bottom panel displays the differential quantities $\Delta A\mathrm{(Li)}$ versus $\Delta$\Teff, defined as the difference between the primary and secondary components of each pair. Here, symbols are color-coded by the \Teff\ of the secondary star and their size scales with $|\Delta v\sin i|$.

It is difficult to find the influences of [Fe/H] on {\it A}(Li) due to the co-natal nature between two components of an individual WB and the narrow range (-0.4$<$[Fe/H]$<$0.2\,dex) of metallicity in our sample. Therefore, in this section, we systematically explore the effects of effective temperature (\Teff) and stellar rotation ($v\sin i$) on the observed Li abundances.

\subsection{{\Teff}: A Key Controlling Factor for Lithium Depletion}
As Figure~\ref{fig:offset2} shows, our data reveal a strong correlation between {\it A}(Li) and \Teff, characterized by three distinct regimes: the Li dip (6200–6600\,K), the Li plateau (6000–6200\,K), and a clear linear trend in the cooler binaries (\Teff $<$ 6000\,K)—where the connecting lines generally share a consistent slope.

\subsubsection{Li Dip (6200\,K -- 6600\,K)}
\label{sec:Lidip}
The upper panel of Figure~\ref{fig:offset2} shows a pronounced Li dip feature in the \Teff\ range 6200--6600\,K. Within binary systems, Li depletion becomes increasingly severe toward the center of this range, reaching a minimum near 6500\,K. This morphology—lower {\it A}(Li) in the center and higher values toward both hotter and cooler edges—is consistent with the classical Li dip observed in intermediate age and old open clusters. 

In particular, a binary system in our sample exhibits an intriguing deviation: both components (connected by a red dashed line) have \Teff $\sim$ 6500\,K, a temperature near the bottom of the Li dip valley where the lowest {\it A}(Li) values are expected. In contrast to this expectation, the primary star shows unexpectedly high Li abundance. The values of its astrometry, atmospheric parameters, and Li abundance are given as WB01 in the first row of Table~\ref{tab:table1}. We will discuss this exceptional case further in Section~\ref{sec:DISCUSSION}.

In the bottom panel of Figure~\ref{fig:offset2}, binaries whose components fall within the Li dip are marked by inverted triangles, color-coded by \Teff\ of the secondary star. The differential abundance of Li, $\Delta A\mathrm{(Li)}$, becomes more pronounced near the center of the dip (around $\sim$6450\,K). In our sample, the absolute value of $\Delta A\mathrm{(Li)}$ within the dip reaches a maximum of 1.65\,dex, which rapidly decreases to 0.97\,dex toward the edges of the temperature range.

\subsubsection{Li Plateau (6000\,K -- 6200\,K)}
\label{sec:Listeep}
The Li plateau appears as another distinct feature in the \Teff\ range of 6000--6200\,K, as shown in the inset of the upper panel of Figure~\ref{fig:offset2}. To highlight this regime, the inset displays only binary pairs in which both components have \Teff\ between 5700\,K and 6200\,K. Across the plateau, {\it A}(Li) remains roughly constant at an average value of about 2.5\,dex, marked by the red dashed horizontal line spanning 5700--6200\,K. This observed average is slightly lower than the theoretically predicted primordial Li abundance of {\it A}(Li) $\sim$ 2.7\,dex from standard Big Bang nucleosynthesis. In the bottom panel of Figure~\ref{fig:offset2}, these binary pairs are indicated by filled squares.

Notably, as shown in the inset in Figure~\ref{fig:offset2}, three binary systems in our sample exhibit slight deviations from the plateau. (1) In binary system WB02, the secondary star is a typical Li-plateau source with \textit{A}(Li) $\simeq$ 2.46\,dex, while the primary star shows a significantly lower abundance (\textit{A}(Li) $\simeq$ 1.49\,dex). This deviation likely occurs because the primary's \Teff\ (6188\,K) is near the Li-dip zone, causing it to exhibit Li-dip characteristics. (2) In binary system WB03, the secondary star (\Teff $\simeq$ 6017\,K) lies close to the lower boundary (6000\,K) of the Li plateau. Its Li abundance (\textit{A}(Li) $\simeq$ 1.03\,dex) is approximately 1.44\,dex lower than that of the primary star. (3) In the WB04 binary system, both components deviate from the Li plateau, and the reason for this deviation remains unclear.

According to \citet{Sun2024}, {\it A}(Li) shows a rapid decrease within $\pm100$\,K around the solar \Teff\ ($\sim$5777\,K). In our sample (upper panel of Figure~\ref{fig:offset2}), over this solar-like range (5677--5877\,K), aside from a pronounced sharp drop for the two binary pairs highlighted by the orange line in the inset, no statistically significant steep decline is observed overall.

\subsubsection{Cool Stars (\textless 6000\,K)}
\label{sec:coolstars}
A clear linear trend is observed for cooler binaries with \Teff$<$ 6000\,K, as shown in the upper panel of Figure~\ref{fig:offset2}. The connecting lines exhibit a consistent slope across a wide $\Delta$\Teff\ range, exceeding 1200\,K in some cases (bottom panel).The dashed black line represents a fit to the filled points (cool stars only), excluding symbols for Li dip binaries (inverted triangles), Li plateau binaries (filled squares), and systems where only the secondary is a cool star (crosses). The fitted slope of $\sim$0.0015 implies that for cool stars, a \Teff\ increase of 100\,K corresponds to an increase of {\it A}(Li) of approximately 0.15\,dex, regardless of the binaries' ages and metallicities.  Although open clusters of different ages show significant scatter in the \Teff\ plane due to continued main-sequence depletion
\citep{Steinhauer2025}, the relatively tight correlation in Figure~\ref{fig:offset2} can be understood by considering the age distribution of our sample. Our wide binaries are field stars, with ages typically exceeding those of most open clusters. In such old G and K dwarfs, lithium depletion has effectively saturated, reducing the dispersion at a given \Teff. Thus, the tight trend reflects the convergence of lithium abundances in old stars, consistent with the idea that \Teff\ (via convection zone depth) is the primary parameter governing the final lithium abundance after several Gyr of evolution.

According to Standard Stellar Evolution Theory, a star's effective temperature dictates the depth of its SCZ. Cooler stars possess deeper SCZs because of their higher atmospheric opacity. A deeper SCZ enhances internal mixing, efficiently transporting Li from the surface to the hot interior, where it is destroyed. This process intensifies post-main-sequence. As the core contracts and the envelope expands, the star cools. The lower temperature increases the opacity (e.g., via neutral H and molecules such as TiO), forcing the outer layers to become more convective. 
The correlation between deeper SCZs in cooler stars and enhanced Li depletion provides a theoretical basis for the observed depletion trend in our cool star sample.

\subsection{$v\sin i$: Is A Dominant Driver for Lithium Depletion?}

Figure~\ref{fig:offset2} (upper panel) shows that stars hotter than 6000\,K generally have higher projected rotational velocities ($v\sin i$) and higher average Li abundances ({\it A}(Li)) than cooler stars. Specifically, hotter stars (\Teff$>$6200\,K) show $\langle v\sin i\rangle\sim 25$\,\kms, while cooler stars (\Teff$<6000$\,K) have $\langle v\sin i\rangle\sim 5$\,\kms. Although this may suggest a link between rotation and Li depletion, our results indicate that the predominant effects are effective temperature.

We find no evidence in our sample that $v\sin i$ independently controls Li depletion. This conclusion is supported by three key observational results:

\begin{enumerate}
    \item \textbf{The correlation is secondary compared to \Teff:} Both $v\sin i$ and {\it A}(Li) scale with stellar mass, and hence \Teff. Hotter stars naturally rotate faster and retain more Li as a result of shallower convective zones. The apparent connection between higher $v\sin i$ and higher {\it A}(Li) is therefore a by-product of their common dependence on temperature.

    \item \textbf{No clear relationship within individual binary pairs:} Even in the Li dip region, we find no systematic trend between $v\sin i$ and {\it A}(Li) within coeval pairs. Several systems show that the component with \textit{higher} $v\sin i$ has \textit{lower} \textit{A}(Li), contrary to the expectation that faster rotators should retain more Li.

    \item \textbf{No correlation when \Teff\ is fixed:} For binary pairs with negligible \Teff\ differences ($\Delta$\Teff$<100$\,K, enclosed by the green dashed box on the bottom panel), the mean $\Delta$\textit{A}(Li) is $\sim 0.03\pm 0.08$\,dex. This holds even for the three pairs with large $\Delta$\text{A}(Li) (marked with inverted triangles and squares), indicating that rotation does not lead to the observed lithium scatter.
\end{enumerate}

These findings do not confirm the claim of \citet{Sun2025b} that faster rotators systematically retain higher \textit{A}(Li) with less scatter. Instead, within our sample, the available evidence does not support $v\sin i$ as an independent dominant driver of the observed lithium depletion, and the dominant control appears to be \Teff.

\section{DISCUSSION}
\label{sec:DISCUSSION}

As mentioned in Section \ref{sec:Lidip}, we identify an interesting binary system (WB01, connected by a red dashed line in Figure~\ref{fig:offset2}). Its primary star exhibits an unexpectedly high Li abundance ($\sim$2.84\,dex) near the bottom of the Li dip valley, approximately $\sim$1.34\,dex greater than that of its secondary companion with nearly identical \Teff. This significant discrepancy warrants further exploration of possible underlying mechanisms. We discuss two plausible scenarios that may account for the observed Li enhancement in the primary star.

One possible explanation is external enrichment via planetesimal accretion or engulfment of hydrogen-poor planets. The elevated Li abundance of the primary star could result from the accretion of external material rich in Li. Two related processes are considered: \textit{planetesimal accretion} and \textit{engulfment of hydrogen-poor planets}. In both cases, Li preserved in solid or planetary bodies is delivered to the stellar surface, altering its observed composition. 
Consequently, even a modest amount of accreted hydrogen-poor material can significantly enhance the surface abundances of Li and other elements. Planetesimal accretion involves the continuous or episodic accumulation of small solid bodies (e.g., asteroids or comet cores), while planetary engulfment is a more catastrophic event where an entire planet is assimilated. 
The observed Li excess in the primary star could thus reflect past accretion or engulfment events, possibly during the system's early evolutionary stages. Further spectroscopic analysis of the ratios of refractory versus volatile elements could help distinguish between these scenarios \citep[e.g.,][]{Nagar_2020}.

\begin{figure}[htbp]
    \centering
    \includegraphics[width=1.
    \linewidth, trim=.5cm 0.5cm 0.3cm 0.3cm, clip]{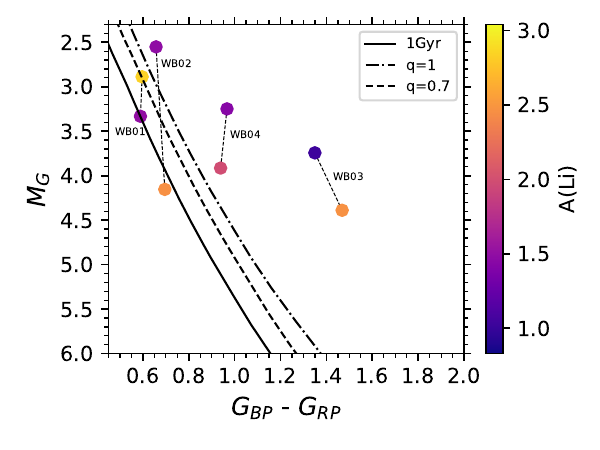}
    \caption{The Color-Magnitude Diagram (CMD) pattern of the anomalous systems (e.g., WB01, WB02, WB03 and WB04) that color-coded by their {\it A}(Li) and connected by the dashed gray lines for the primary and secondary stars. The black solid line represents the 1\,Gyr MS isochrone, and the dashed and dashed-dot black lines are the isochrones for unresolved binaries with q = 0.7 and 1.0, respectively. The q represent the mass ratio of an unresolved binary.}
    \label{fig:unresovled_comp}
\end{figure}

Another plausible explanation involves the presence of an unresolved companion and binary interaction effects. Photometric and isochronal analysis supports this: as shown in Figure~\ref{fig:unresovled_comp}, the secondary star aligns with a 1\,Gyr isochrone (black solid line), while the primary star's position is consistent with an unresolved binary system having a mass ratio of $q = 0.7$ (the dashed black isochrone calculated according to \citet{LiLu2020ApJ}). This implies that the primary may have an unresolved companion of about $0.7\,\rm M_{primary}$. The unresolved companion could influence the primary's Li abundance through several binary interaction pathways. For instance, tidal interactions in close binary systems can induce rotational mixing, enhancing Li depletion or production, depending on the specific angular momentum transport. Alternatively, mass transfer from the companion to the primary—either via Roche-lobe overflow or wind accretion—could deposit Li-rich material onto the primary's surface. Such processes are known to alter the surface compositions in binary systems \citep[e.g., in short-period tidally locked binaries;][]{Deliyannis1994ApJ}.  
However, the specific physical process through which an unresolved companion might enhance the primary's Li abundance remains unclear and requires detailed modeling of binary evolution and internal mixing to elucidate. Additionally, the observed Li line strength itself could be affected by spectral contamination from the light of an unresolved companion, further complicating the interpretation. It is also important to note that this particular scenario—invoking an unresolved companion to explain Li enhancement—cannot account for other binary systems in our sample (e.g., WB02, WB03, and WB04). In those systems, the primary stars exhibit low Li abundances despite potentially hosting unresolved companions, indicating that the mere presence of a close companion does not systematically lead to Li enrichment.

In summary, the anomalous Li abundance in the primary star of this binary system can be attributed either to external enrichment (through planetesimal accretion or planetary engulfment) or to binary interactions involving an unresolved companion. Both hypotheses are plausible and align with broader findings in stellar astrophysics, particularly for A-type stars, where surface abundances are sensitive to external material and binary dynamics. Future high-resolution spectroscopic observations, coupled with detailed chemical abundance analysis and binary evolution modeling, will be essential to discriminate between these scenarios and uncover the origin of Li enhancement in such systems.

\section{CONCLUSION}
\label{sec:CONCLUSION}

This study utilizes a sample of 116 wide binary systems as coeval and chemically homogeneous stellar pairs to investigate the factors governing Li depletion in main-sequence stars. The primary conclusions drawn from our analysis are as follows:

1. \textbf{Effective temperature (\Teff) is the dominant factor controlling Li depletion:} (a) The well-established morphology of the Li–\Teff\ relation is clearly recovered in our binary sample. The infiltration is most severe near its center around 6500\,K; (b) the Li Plateau (6000–6200\,K), where the abundances remain roughly constant at an average \textit{A}(Li) $\sim$2.5\,dex; and (c) cooler stars (\Teff $<$ 6000\,K), where \textit{A}(Li) exhibits a clear linear correlation with \Teff, increasing by $\sim$0.15\,dex per 100\,K increase. The consistent slopes of the connecting lines between binary components in a wide \Teff\ range reinforce that the depletion trend is primarily governed by temperature, likely through its control on the depth of SCZ and the efficiency of internal mixing.

2. \textbf{Projected rotational velocity ($v\sin i$) is not an independent dominant driver of Li depletion in our sample:} Although hotter ($>$6000\,K), faster-rotating stars show higher average Li abundances, this correlation is secondary to the underlying dependence on \Teff. No systematic relationship between $v\sin i$ and \textit{A}(Li) is found within individual binary pairs of similar temperature, nor is a correlation present when the differences in \Teff\ are minimal ($\Delta$\Teff $<100$\,K). Therefore, the observed link between rotation and Li abundances can best be explained by their common scaling with stellar mass and effective temperature, rather than a direct causal influence.

3. \textbf{An exceptional binary system within the Li dip presents a case for anomalous Li enrichment:} One system, with both components with \Teff\ near 6500\,K, shows a $\sim$1.4\,dex Li excess in its primary star compared to its secondary. Two plausible non-mutually exclusive origins are suggested for this anomaly: (i) external enrichment via accretion of planetesimals or engulfment of a hydrogen-poor planet, or (ii) binary interactions with an unresolved tertiary companion that could induce rotational mixing or mass transfer. Both scenarios are observationally motivated but require further verification through detailed chemical abundance analysis and binary evolution modeling.

In summary, our analysis of wide binaries confirms that stellar effective temperature is the principal parameter governing the observed pattern of LI depletion across a broad spectral range. The role of rotation appears to be subsidiary in this context. The identified anomalous system highlights that additional processes—such as external pollution or complex binary interactions—can occasionally override the standard depletion trends and merit targeted follow-up. Future studies combining high-resolution spectroscopy for detailed elemental abundances and precise orbital monitoring will be crucial to elucidate these exceptional cases and further refine our understanding of lithium evolution in stellar systems.

\begin{acknowledgments}
The authors thank Min Fang, Xiao-Ting Fu, Feng Wang, Lu Li and Xun-Zhou Chen for the helpful discussions. This work is supported by the Strategic Priority Research Program of Chinese Academy of Sciences, grant No.\,XDB1160101. This research is also supported by the National Natural Science Foundation of China under Grant Nos.\,12373033, 12090040/4, 12373033, 12373036, 12222305, 12427804, the National Key R$\rm\& $D Program of China 2024YFA1611903, the Scientific Instrument Developing Project of the Chinese Academy of Sciences, Grant No.\,ZDKYYQ20220009,  the Key Project of Zhejiang Provincial Natural Science Foundation (No.\,ZCLZ25A0301),and the International Partnership Program of the Chinese Academy of Sciences, Grant No.\,178GJHZ2022047GC. 

\end{acknowledgments}

\bibliography{bib}
\bibliographystyle{aasjournal}
\appendix
Table \ref{tab:table1} lists the stellar parameters of the binaries used in this study, including the coordinates, astrometry, atmospheric parameters and Li abundances. The complete version of this table is available in the online Journal in the machine-readable table format.

\input{tableA1}



\end{document}

%% file: tableA1.tex
\begin{deluxetable*}{ccccccccccccccccc}
\renewcommand{\thetable}{A\arabic{table}}
\setcounter{table}{0}
\caption{The Coordinates, Astrometry, Atmospheric Parameters and Li Abundances of the Binaries}
\label{tab:table1}
\tablehead{
\colhead{Name} & \colhead{GALAH ID} & \colhead{RA} & \colhead{DEC} & \colhead{$\mu_{\alpha_*}$} & \colhead{$\mu_{\delta}$} & \colhead{$\pi$} & \colhead{$\rm M_G$} & \colhead{$\rm G_{BP-RP}$} & \colhead{\Teff} & \colhead{log $g$} & \colhead{[Fe/H]} & \colhead{$\xi_{\rm t}$} &  \colhead{$R_V$} & \colhead{$v\sin i$} &
\colhead{A(Li)}\\
\colhead{} & \colhead{} & \multicolumn{2}{c}{(deg)}  & 
\multicolumn{2}{c}{(mas/yr)} & 
\colhead{(mas)} & 
\multicolumn{2}{c}{(mag)} & \colhead{(K)} & 
\colhead{} & 
\colhead{(dex)} &
\multicolumn{3}{c}{(km~s$^{-1}$)} & 
\colhead{(dex)}}

\startdata
WB01a & 161212001601321 & 74.01 & -62.05 & -3.78$\pm$0.01 & -6.88$\pm$0.02 & 1.17$\pm$0.01 & 2.89 & 0.60 & 6431$\pm$82 & 4.0$\pm$0.2 & -0.20$\pm$0.08 & 1.5 & 4.0$\pm$0.5 & 13.0$\pm$2.3 & 2.84 \\
WB01b & 171104002301321 & 74.01 & -62.05 & -3.92$\pm$0.02 & -6.89$\pm$0.02 & 1.19$\pm$0.01 & 3.33 & 0.59 & 6438$\pm$84 & 4.1$\pm$0.2 & -0.20$\pm$0.09 & 1.5 & 5.1$\pm$0.6 & 16.0$\pm$2.3 & 1.50 \\
WB02a & 161217002601243 & 65.66 & -60.09 & 2.41$\pm$0.01 & -6.73$\pm$0.01 & 2.06$\pm$0.01 & 2.55 & 0.66 & 6188$\pm$74 & 3.9$\pm$0.2 & -0.21$\pm$0.05 & 1.4 & 23.7$\pm$0.4 & 10.0$\pm$2.1 & 1.49 \\
WB02b & 161217002601252 & 65.71 & -60.10 & 2.41$\pm$0.01 & -6.72$\pm$0.01 & 2.06$\pm$0.01 & 4.15 & 0.70 & 6107$\pm$96 & 4.3$\pm$0.2 & -0.17$\pm$0.10 & 1.3 & 23.9$\pm$0.6 & 6.9$\pm$2.4 & 2.46 \\
WB03a & 150411006101052 & 249.46 & -19.92 & -6.21$\pm$0.02 & -9.54$\pm$0.02 & 1.80$\pm$0.02 & 3.74 & 1.35 & 6017$\pm$84 & 3.7$\pm$0.2 & -0.01$\pm$0.07 & 1.3 & -22.5$\pm$0.4 & 7.7$\pm$2.2 & 1.03 \\
WB03b & 150411006101048 & 249.48 & -19.90 & -6.29$\pm$0.06 & -9.61$\pm$0.04 & 1.87$\pm$0.05 & 4.39 & 1.47 & 6180$\pm$90 & 3.9$\pm$0.2 & -0.10$\pm$0.09 & 1.4 & -22.1$\pm$0.5 & 9.0$\pm$2.3 & 2.47 \\
WB04a & 181226002601276 & 69.92 & 3.99 & -2.01$\pm$0.02 & -4.33$\pm$0.01 & 1.24$\pm$0.02 & 3.25 & 0.97 & 6094$\pm$79 & 3.9$\pm$0.2 & 0.11$\pm$0.06 & 1.3 & 16.4$\pm$0.4 & 10.3$\pm$2.2 & 1.44 \\
WB04b & 181226002601213 & 69.91 & 3.98 & -2.07$\pm$0.02 & -4.38$\pm$0.01 & 1.26$\pm$0.02 & 3.91 & 0.94 & 6178$\pm$84 & 4.2$\pm$0.2 & 0.11$\pm$0.07 & 1.4 & 16.3$\pm$0.5 & 12.3$\pm$2.2 & 1.98 \\
\enddata
\tablecomments{The GALAH DR3 identifiers of each star are given in column 2. The coordinates are given in the column 3 and 4. The column 5 to 6 represent the proper motion at RA and DEC. The effective temperatures ($T_{\mathrm{eff}}$), the logarithmic of the surface gravity (log $g$) and the metallicity ([Fe/H]) with their errors in column 10, 11 and 12, respectively. The Microturbulence, the radial velocity, and the project rotational velocity are in the column 13 to 15. The A(Li) we measured are in the last column in this table.}
\end{deluxetable*}

\begin{deluxetable*}{ccccccccccccccccc}
\renewcommand{\thetable}{A\arabic{table}}
\setcounter{table}{0}
\caption{The Coordinates, Astrometry, Atmospheric Parameters and Li Abundances of the Binaries}
\tablehead{
\colhead{Name} & \colhead{GALAH ID} & \colhead{RA} & \colhead{DEC} & \colhead{$\mu_{\alpha_*}$} & \colhead{$\mu_{\delta}$} & \colhead{$\pi$} & \colhead{$\rm M_G$} & \colhead{$\rm G_{BP-RP}$} & \colhead{\Teff} & \colhead{log $g$} & \colhead{[Fe/H]} & \colhead{$\xi_{\rm t}$} &  \colhead{$R_V$} & \colhead{$v\sin i$} &
\colhead{A(Li)}\\
\colhead{} & \colhead{} & \multicolumn{2}{c}{(deg)}  & 
\multicolumn{2}{c}{(mas/yr)} & 
\colhead{(mas)} & 
\multicolumn{2}{c}{(mag)} & \colhead{(K)} & 
\colhead{} & 
\colhead{(dex)} &
\multicolumn{3}{c}{(km~s$^{-1}$)} & 
\colhead{(dex)}}

\startdata
WB15a & 161228002501291 & 127.04 & 17.76 & -4.57$\pm$0.02 & -6.20$\pm$0.01 & 2.53$\pm$0.01 & 4.56 & 0.85 & 5741$\pm$109 & 4.4$\pm$0.2 & 0.17$\pm$0.11 & 1.1 & 45.7$\pm$0.6 & 7.7$\pm$2.5 & 1.68 \\
WB15b & 160110002601291 & 127.04 & 17.77 & -4.50$\pm$0.02 & -6.34$\pm$0.01 & 2.48$\pm$0.02 & 4.96 & 0.91 & 5468$\pm$118 & 4.4$\pm$0.2 & 0.18$\pm$0.12 & 1.1 & 45.4$\pm$0.6 & 7.8$\pm$2.5 & 1.65 \\
WB16a & 161006004901134 & 16.93 & -1.73 & -4.92$\pm$0.02 & -18.02$\pm$0.02 & 1.39$\pm$0.02 & 4.52 & 0.83 & 5796$\pm$97 & 4.3$\pm$0.2 & -0.41$\pm$0.09 & 1.2 & -20.3$\pm$0.6 & 6.6$\pm$2.5 & 2.00 \\
WB16b & 161006004901129 & 16.94 & -1.73 & -5.03$\pm$0.03 & -17.95$\pm$0.02 & 1.28$\pm$0.03 & 5.33 & 0.99 & 5356$\pm$124 & 4.3$\pm$0.2 & -0.39$\pm$0.11 & 1.1 & -19.9$\pm$0.6 & 4.7$\pm$2.7 & 1.40 \\
WB17a & 150827002901326 & 284.14 & -53.67 & 1.69$\pm$0.01 & -13.07$\pm$0.01 & 2.27$\pm$0.01 & 3.91 & 0.71 & 6263$\pm$111 & 4.3$\pm$0.2 & -0.09$\pm$0.13 & 1.4 & -14.9$\pm$0.8 & 10.8$\pm$2.6 & 2.83 \\
WB17b & 160812003101336 & 284.13 & -53.67 & 2.13$\pm$0.01 & -12.92$\pm$0.01 & 2.29$\pm$0.02 & 5.68 & 1.02 & 5298$\pm$104 & 4.5$\pm$0.2 & -0.02$\pm$0.08 & 1.1 & -15.2$\pm$0.5 & 5.8$\pm$2.3 & 1.21 \\
WB18a & 140314002601294 & 109.60 & -64.69 & -13.18$\pm$0.01 & 33.05$\pm$0.01 & 4.23$\pm$0.01 & 5.33 & 0.96 & 5433$\pm$74 & 4.5$\pm$0.2 & 0.20$\pm$0.05 & 1.1 & 9.5$\pm$0.3 & 6.0$\pm$2.1 & 0.95 \\
WB18b & 140314002601297 & 109.66 & -64.70 & -13.14$\pm$0.01 & 33.03$\pm$0.01 & 4.22$\pm$0.01 & 5.44 & 0.98 & 5319$\pm$75 & 4.5$\pm$0.2 & 0.13$\pm$0.05 & 1.1 & 9.5$\pm$0.3 & 5.7$\pm$2.1 & 1.18 \\
\enddata
\end{deluxetable*}

\begin{deluxetable*}{ccccccccccccccccc}
\renewcommand{\thetable}{A\arabic{table}}
\setcounter{table}{0}
\caption{The Coordinates, Astrometry, Atmospheric Parameters and Li Abundances of the Binaries}
\tablehead{
\colhead{Name} & \colhead{GALAH ID} & \colhead{RA} & \colhead{DEC} & \colhead{$\mu_{\alpha_*}$} & \colhead{$\mu_{\delta}$} & \colhead{$\pi$} & \colhead{$\rm M_G$} & \colhead{$\rm G_{BP-RP}$} & \colhead{\Teff} & \colhead{log $g$} & \colhead{[Fe/H]} & \colhead{$\xi_{\rm t}$} &  \colhead{$R_V$} & \colhead{$v\sin i$} &
\colhead{A(Li)}\\
\colhead{} & \colhead{} & \multicolumn{2}{c}{(deg)}  & 
\multicolumn{2}{c}{(mas/yr)} & 
\colhead{(mas)} & 
\multicolumn{2}{c}{(mag)} & \colhead{(K)} & 
\colhead{} & 
\colhead{(dex)} &
\multicolumn{3}{c}{(km~s$^{-1}$)} & 
\colhead{(dex)}}

\startdata
WB30a & 150827002901326 & 284.14 & -53.67 & 1.69$\pm$0.01 & -13.07$\pm$0.01 & 2.27$\pm$0.01 & 3.91 & 0.71 & 6263$\pm$111 & 4.3$\pm$0.2 & -0.09$\pm$0.13 & 1.4 & -14.9$\pm$0.8 & 10.8$\pm$2.6 & 2.83 \\
WB30b & 160812003101336 & 284.13 & -53.67 & 2.13$\pm$0.01 & -12.92$\pm$0.01 & 2.29$\pm$0.02 & 5.68 & 1.02 & 5298$\pm$104 & 4.5$\pm$0.2 & -0.02$\pm$0.08 & 1.1 & -15.2$\pm$0.5 & 5.8$\pm$2.3 & 1.21 \\
WB31a & 140314002601294 & 109.60 & -64.69 & -13.18$\pm$0.01 & 33.05$\pm$0.01 & 4.23$\pm$0.01 & 5.33 & 0.96 & 5433$\pm$74 & 4.5$\pm$0.2 & 0.20$\pm$0.05 & 1.1 & 9.5$\pm$0.3 & 6.0$\pm$2.1 & 0.95 \\
WB31b & 140314002601297 & 109.66 & -64.70 & -13.14$\pm$0.01 & 33.03$\pm$0.01 & 4.22$\pm$0.01 & 5.44 & 0.98 & 5319$\pm$75 & 4.5$\pm$0.2 & 0.13$\pm$0.05 & 1.1 & 9.5$\pm$0.3 & 5.7$\pm$2.1 & 1.18 \\
WB32a & 171102004501355 & 71.21 & 6.11 & -1.94$\pm$0.02 & -3.52$\pm$0.01 & 2.22$\pm$0.01 & 3.83 & 0.78 & 6279$\pm$78 & 4.2$\pm$0.2 & -0.17$\pm$0.06 & 1.4 & 14.8$\pm$0.4 & 9.2$\pm$2.2 & 2.27 \\
WB32b & 171102004501353 & 71.18 & 6.10 & -1.86$\pm$0.02 & -3.42$\pm$0.01 & 2.17$\pm$0.01 & 4.91 & 0.95 & 5931$\pm$116 & 4.5$\pm$0.2 & -0.02$\pm$0.11 & 1.2 & 14.3$\pm$0.7 & 7.2$\pm$2.6 & 2.22 \\
WB33a & 151008004001147 & 65.42 & 18.01 & 7.88$\pm$0.02 & -16.25$\pm$0.02 & 3.00$\pm$0.01 & 5.30 & 1.23 & 5679$\pm$98 & 4.3$\pm$0.2 & -0.05$\pm$0.08 & 1.1 & 47.8$\pm$0.5 & 4.5$\pm$2.4 & 1.75 \\
WB33b & 151008004001149 & 65.41 & 18.01 & 7.89$\pm$0.02 & -16.34$\pm$0.01 & 3.00$\pm$0.01 & 5.30 & 1.21 & 5698$\pm$98 & 4.3$\pm$0.2 & 0.04$\pm$0.09 & 1.1 & 47.7$\pm$0.5 & 6.6$\pm$2.4 & 2.01 \\
\enddata
\end{deluxetable*}

\begin{deluxetable*}{ccccccccccccccccc}
\renewcommand{\thetable}{A\arabic{table}}
\setcounter{table}{0}
\caption{The Coordinates, Astrometry, Atmospheric Parameters and Li Abundances of the Binaries}
\tablehead{
\colhead{Name} & \colhead{GALAH ID} & \colhead{RA} & \colhead{DEC} & \colhead{$\mu_{\alpha_*}$} & \colhead{$\mu_{\delta}$} & \colhead{$\pi$} & \colhead{$\rm M_G$} & \colhead{$\rm G_{BP-RP}$} & \colhead{\Teff} & \colhead{log $g$} & \colhead{[Fe/H]} & \colhead{$\xi_{\rm t}$} &  \colhead{$R_V$} & \colhead{$v\sin i$} &
\colhead{A(Li)}\\
\colhead{} & \colhead{} & \multicolumn{2}{c}{(deg)}  & 
\multicolumn{2}{c}{(mas/yr)} & 
\colhead{(mas)} & 
\multicolumn{2}{c}{(mag)} & \colhead{(K)} & 
\colhead{} & 
\colhead{(dex)} &
\multicolumn{3}{c}{(km~s$^{-1}$)} & 
\colhead{(dex)}}

\startdata
WB45a & 141104002301335 & 338.53 & -13.25 & -7.82$\pm$0.02 & -21.61$\pm$0.02 & 4.08$\pm$0.02 & 6.09 & 1.25 & 4967$\pm$88 & 4.6$\pm$0.2 & 0.41$\pm$0.07 & 1 & -14.2$\pm$0.4 & 6.4$\pm$2.2 & 0.80 \\
WB45b & 141104002301275 & 338.52 & -13.25 & -7.79$\pm$0.02 & -21.90$\pm$0.02 & 4.11$\pm$0.02 & 6.60 & 1.41 & 4717$\pm$97 & 4.6$\pm$0.2 & 0.37$\pm$0.09 & 1 & -14.1$\pm$0.5 & 5.7$\pm$2.3 & 0.49 \\
WB46a & 141104002301070 & 338.95 & -14.11 & 1.25$\pm$0.02 & 5.12$\pm$0.01 & 3.43$\pm$0.02 & 5.22 & 0.89 & 5496$\pm$85 & 4.5$\pm$0.2 & -0.11$\pm$0.06 & 1.1 & 7.6$\pm$0.4 & 7.7$\pm$2.2 & 1.88 \\
WB46b & 141104002301072 & 338.93 & -14.12 & 1.24$\pm$0.02 & 5.15$\pm$0.02 & 3.45$\pm$0.02 & 6.52 & 1.23 & 4747$\pm$107 & 4.6$\pm$0.2 & -0.06$\pm$0.09 & 1 & 7.5$\pm$0.5 & 6.9$\pm$2.4 & 0.65 \\
WB47a & 140711003401166 & 330.76 & -8.45 & 62.22$\pm$0.02 & 0.71$\pm$0.01 & 6.15$\pm$0.02 & 6.47 & 1.28 & 4765$\pm$77 & 4.6$\pm$0.2 & 0.19$\pm$0.05 & 1 & -23.3$\pm$0.3 & 5.6$\pm$2.1 & 0.04 \\
WB47b & 140711003401005 & 330.79 & -8.41 & 62.40$\pm$0.02 & 0.87$\pm$0.02 & 6.16$\pm$0.02 & 6.73 & 1.36 & 4635$\pm$81 & 4.6$\pm$0.2 & 0.14$\pm$0.06 & 1 & -23.2$\pm$0.4 & 6.1$\pm$2.2 & 0.01 \\
WB48a & 140708005301013 & 333.84 & -6.85 & 3.39$\pm$0.02 & 12.90$\pm$0.02 & 1.17$\pm$0.02 & 3.46 & 1.06 & 5061$\pm$182 & 3.7$\pm$0.2 & -0.15$\pm$0.14 & 1.1 & -42.5$\pm$0.7 & 7.3$\pm$2.6 & 1.45 \\
WB48b & 151009002601014 & 333.84 & -6.86 & 3.37$\pm$0.02 & 13.03$\pm$0.02 & 1.19$\pm$0.02 & 3.72 & 0.80 & 5845$\pm$105 & 4.0$\pm$0.2 & -0.14$\pm$0.10 & 1.2 & -43.0$\pm$0.6 & 6.0$\pm$2.5 & 2.25 \\
\enddata
\end{deluxetable*}

\begin{deluxetable*}{ccccccccccccccccc}
\renewcommand{\thetable}{A\arabic{table}}
\setcounter{table}{0}
\caption{The Coordinates, Astrometry, Atmospheric Parameters and Li Abundances of the Binaries}
\tablehead{
\colhead{Name} & \colhead{GALAH ID} & \colhead{RA} & \colhead{DEC} & \colhead{$\mu_{\alpha_*}$} & \colhead{$\mu_{\delta}$} & \colhead{$\pi$} & \colhead{$\rm M_G$} & \colhead{$\rm G_{BP-RP}$} & \colhead{\Teff} & \colhead{log $g$} & \colhead{[Fe/H]} & \colhead{$\xi_{\rm t}$} &  \colhead{$R_V$} & \colhead{$v\sin i$} &
\colhead{A(Li)}\\
\colhead{} & \colhead{} & \multicolumn{2}{c}{(deg)}  & 
\multicolumn{2}{c}{(mas/yr)} & 
\colhead{(mas)} & 
\multicolumn{2}{c}{(mag)} & \colhead{(K)} & 
\colhead{} & 
\colhead{(dex)} &
\multicolumn{3}{c}{(km~s$^{-1}$)} & 
\colhead{(dex)}}

\startdata
WB60a & 170216003301017 & 190.83 & -29.69 & 30.67$\pm$0.02 & -5.65$\pm$0.01 & 1.80$\pm$0.02 & 4.87 & 0.90 & 5465$\pm$101 & 4.3$\pm$0.2 & -0.29$\pm$0.08 & 1.1 & 0.1$\pm$0.5 & 4.9$\pm$2.4 & 1.34 \\
WB60b & 170216003301016 & 190.84 & -29.68 & 30.66$\pm$0.02 & -5.71$\pm$0.01 & 1.81$\pm$0.02 & 5.03 & 0.94 & 5386$\pm$104 & 4.4$\pm$0.2 & -0.18$\pm$0.09 & 1.1 & -0.3$\pm$0.5 & 4.5$\pm$2.4 & 1.30 \\
WB61a & 151110004201385 & 110.24 & 8.89 & 1.56$\pm$0.02 & -4.90$\pm$0.02 & 1.82$\pm$0.02 & 2.15 & 0.58 & 6518$\pm$86 & 3.8$\pm$0.2 & -0.10$\pm$0.10 & 1.6 & -13.8$\pm$0.6 & 11.9$\pm$2.3 & 2.76 \\
WB61b & 151110004201386 & 110.23 & 8.87 & 1.60$\pm$0.02 & -5.01$\pm$0.02 & 1.80$\pm$0.02 & 2.51 & 0.54 & 6630$\pm$101 & 4.0$\pm$0.2 & -0.01$\pm$0.14 & 1.7 & -22.6$\pm$1.3 & 23.1$\pm$2.8 & 3.03 \\
WB62a & 151110004201055 & 109.92 & 8.04 & -38.11$\pm$0.02 & -51.70$\pm$0.01 & 4.08$\pm$0.01 & 3.62 & 0.71 & 5991$\pm$79 & 4.1$\pm$0.2 & -0.41$\pm$0.06 & 1.2 & -30.2$\pm$0.4 & 5.1$\pm$2.2 & 2.38 \\
WB62b & 151110004201051 & 109.91 & 8.06 & -38.22$\pm$0.02 & -51.94$\pm$0.01 & 4.11$\pm$0.01 & 4.12 & 0.71 & 5980$\pm$88 & 4.2$\pm$0.2 & -0.43$\pm$0.08 & 1.2 & -29.4$\pm$0.5 & 6.3$\pm$2.3 & 2.15 \\
WB63a & 170506005401269 & 286.49 & -20.23 & 5.57$\pm$0.02 & 0.88$\pm$0.02 & 2.23$\pm$0.02 & 2.83 & 0.72 & 6492$\pm$77 & 3.9$\pm$0.2 & -0.15$\pm$0.08 & 1.6 & -19.7$\pm$0.9 & 36.2$\pm$2.5 & 1.67 \\
WB63b & 170506005401263 & 286.47 & -20.23 & 5.60$\pm$0.02 & 0.80$\pm$0.01 & 2.22$\pm$0.02 & 2.84 & 0.73 & 6537$\pm$77 & 4.0$\pm$0.2 & -0.11$\pm$0.08 & 1.6 & -20.1$\pm$0.9 & 35.7$\pm$2.5 & 2.45 \\
\enddata
\end{deluxetable*}

\begin{deluxetable*}{ccccccccccccccccc}
\renewcommand{\thetable}{A\arabic{table}}
\setcounter{table}{0}
\caption{The Coordinates, Astrometry, Atmospheric Parameters and Li Abundances of the Binaries}
\tablehead{
\colhead{Name} & \colhead{GALAH ID} & \colhead{RA} & \colhead{DEC} & \colhead{$\mu_{\alpha_*}$} & \colhead{$\mu_{\delta}$} & \colhead{$\pi$} & \colhead{$\rm M_G$} & \colhead{$\rm G_{BP-RP}$} & \colhead{\Teff} & \colhead{log $g$} & \colhead{[Fe/H]} & \colhead{$\xi_{\rm t}$} &  \colhead{$R_V$} & \colhead{$v\sin i$} &
\colhead{A(Li)}\\
\colhead{} & \colhead{} & \multicolumn{2}{c}{(deg)}  & 
\multicolumn{2}{c}{(mas/yr)} & 
\colhead{(mas)} & 
\multicolumn{2}{c}{(mag)} & \colhead{(K)} & 
\colhead{} & 
\colhead{(dex)} &
\multicolumn{3}{c}{(km~s$^{-1}$)} & 
\colhead{(dex)}}

\startdata
WB75a & 151009004601070 & 61.69 & 13.26 & 11.90$\pm$0.02 & -2.03$\pm$0.01 & 1.79$\pm$0.02 & 4.17 & 1.27 & 5803$\pm$112 & 3.8$\pm$0.2 & -0.50$\pm$0.13 & 1.2 & 36.3$\pm$0.7 & 6.7$\pm$2.7 & 2.20 \\
WB75b & 160106002101085 & 61.69 & 13.26 & 11.97$\pm$0.03 & -1.94$\pm$0.02 & 1.81$\pm$0.02 & 5.51 & 1.27 & 5735$\pm$124 & 4.3$\pm$0.2 & -0.59$\pm$0.15 & 1.1 & 37.1$\pm$0.8 & 5.7$\pm$2.9 & 1.95 \\
WB76a & 160514003301047 & 228.12 & -18.56 & -20.97$\pm$0.02 & -7.94$\pm$0.02 & 2.14$\pm$0.02 & 3.72 & 0.82 & 5916$\pm$76 & 4.1$\pm$0.2 & -0.24$\pm$0.05 & 1.2 & -2.0$\pm$0.4 & 5.9$\pm$2.2 & 2.32 \\
WB76b & 160514003301136 & 228.12 & -18.54 & -20.96$\pm$0.02 & -7.89$\pm$0.03 & 2.12$\pm$0.02 & 4.10 & 0.81 & 5859$\pm$85 & 4.2$\pm$0.2 & -0.37$\pm$0.07 & 1.2 & -1.9$\pm$0.4 & 6.1$\pm$2.3 & 2.08 \\
WB77a & 190207002701019 & 171.83 & 1.65 & -7.31$\pm$0.02 & -16.59$\pm$0.01 & 2.43$\pm$0.02 & 4.82 & 0.81 & 5661$\pm$97 & 4.4$\pm$0.2 & -0.25$\pm$0.08 & 1.1 & 38.6$\pm$0.5 & 5.7$\pm$2.4 & 1.78 \\
WB77b & 190207002701018 & 171.84 & 1.66 & -7.22$\pm$0.02 & -16.61$\pm$0.01 & 2.43$\pm$0.02 & 4.94 & 0.84 & 5544$\pm$104 & 4.4$\pm$0.2 & -0.21$\pm$0.09 & 1.1 & 38.9$\pm$0.5 & 5.6$\pm$2.4 & 1.91 \\
WB78a & 150409005601005 & 174.20 & 1.38 & 4.29$\pm$0.02 & 3.11$\pm$0.02 & 1.66$\pm$0.02 & 3.60 & 0.63 & 6342$\pm$88 & 4.2$\pm$0.2 & -0.23$\pm$0.09 & 1.5 & -5.5$\pm$0.6 & 13.5$\pm$2.3 & 2.21 \\
WB78b & 150427002301018 & 174.20 & 1.38 & 3.99$\pm$0.02 & 2.96$\pm$0.02 & 1.63$\pm$0.02 & 4.12 & 0.71 & 6056$\pm$129 & 4.3$\pm$0.2 & -0.24$\pm$0.17 & 1.3 & -5.7$\pm$1.0 & 8.7$\pm$3.0 & 2.34 \\
\enddata
\end{deluxetable*}

\begin{deluxetable*}{ccccccccccccccccc}
\renewcommand{\thetable}{A\arabic{table}}
\setcounter{table}{0}
\caption{The Coordinates, Astrometry, Atmospheric Parameters and Li Abundances of the Binaries}
\tablehead{
\colhead{Name} & \colhead{GALAH ID} & \colhead{RA} & \colhead{DEC} & \colhead{$\mu_{\alpha_*}$} & \colhead{$\mu_{\delta}$} & \colhead{$\pi$} & \colhead{$\rm M_G$} & \colhead{$\rm G_{BP-RP}$} & \colhead{\Teff} & \colhead{log $g$} & \colhead{[Fe/H]} & \colhead{$\xi_{\rm t}$} &  \colhead{$R_V$} & \colhead{$v\sin i$} &
\colhead{A(Li)}\\
\colhead{} & \colhead{} & \multicolumn{2}{c}{(deg)}  & 
\multicolumn{2}{c}{(mas/yr)} & 
\colhead{(mas)} & 
\multicolumn{2}{c}{(mag)} & \colhead{(K)} & 
\colhead{} & 
\colhead{(dex)} &
\multicolumn{3}{c}{(km~s$^{-1}$)} & 
\colhead{(dex)}}

\startdata
WB90a & 161011003401245 & 65.47 & -64.75 & 4.47$\pm$0.01 & 8.47$\pm$0.02 & 1.18$\pm$0.01 & 2.69 & 0.58 & 6514$\pm$76 & 4.0$\pm$0.2 & -0.24$\pm$0.07 & 1.6 & 18.7$\pm$0.7 & 33.9$\pm$2.3 & 1.97 \\
WB90b & 161011003401246 & 65.40 & -64.75 & 4.56$\pm$0.01 & 8.51$\pm$0.02 & 1.18$\pm$0.01 & 3.19 & 0.58 & 6543$\pm$81 & 4.1$\pm$0.2 & -0.20$\pm$0.08 & 1.6 & 19.7$\pm$0.5 & 10.9$\pm$2.2 & 2.25 \\
WB91a & 161213003101153 & 65.89 & -64.71 & 27.53$\pm$0.01 & 23.18$\pm$0.02 & 2.96$\pm$0.01 & 3.94 & 0.80 & 5877$\pm$76 & 4.2$\pm$0.2 & 0.20$\pm$0.05 & 1.2 & 50.1$\pm$0.3 & 5.8$\pm$2.1 & 2.05 \\
WB91b & 161011003401266 & 65.90 & -64.71 & 27.42$\pm$0.01 & 23.05$\pm$0.02 & 2.98$\pm$0.01 & 4.40 & 0.85 & 5766$\pm$73 & 4.3$\pm$0.2 & 0.20$\pm$0.05 & 1.2 & 50.5$\pm$0.3 & 6.3$\pm$2.1 & 1.15 \\
WB92a & 181221001601191 & 55.96 & -68.57 & 45.75$\pm$0.01 & 75.13$\pm$0.01 & 8.61$\pm$0.01 & 6.07 & 1.17 & 4944$\pm$80 & 4.6$\pm$0.2 & 0.33$\pm$0.06 & 1 & 14.5$\pm$0.4 & 4.4$\pm$2.2 & 0.68 \\
WB92b & 181221001601195 & 56.02 & -68.51 & 45.86$\pm$0.01 & 74.93$\pm$0.01 & 8.64$\pm$0.01 & 6.83 & 1.44 & 4603$\pm$88 & 4.6$\pm$0.2 & 0.24$\pm$0.07 & 1 & 14.8$\pm$0.4 & 4.2$\pm$2.3 & 0.38 \\
WB93a & 171206001602090 & 62.20 & -64.73 & 15.85$\pm$0.01 & 12.18$\pm$0.01 & 2.39$\pm$0.01 & 4.17 & 0.71 & 6092$\pm$70 & 4.3$\pm$0.2 & -0.19$\pm$0.04 & 1.3 & 2.9$\pm$0.3 & 6.8$\pm$2.1 & 2.38 \\
WB93b & 171230001602367 & 62.19 & -64.74 & 15.59$\pm$0.02 & 12.12$\pm$0.02 & 2.43$\pm$0.01 & 4.47 & 0.77 & 5982$\pm$72 & 4.4$\pm$0.2 & -0.19$\pm$0.05 & 1.2 & 2.7$\pm$0.3 & 6.3$\pm$2.1 & 2.46 \\
\enddata
\end{deluxetable*}

\begin{deluxetable*}{ccccccccccccccccc}
\renewcommand{\thetable}{A\arabic{table}}
\setcounter{table}{0}
\caption{The Coordinates, Astrometry, Atmospheric Parameters and Li Abundances of the Binaries}
\tablehead{
\colhead{Name} & \colhead{GALAH ID} & \colhead{RA} & \colhead{DEC} & \colhead{$\mu_{\alpha_*}$} & \colhead{$\mu_{\delta}$} & \colhead{$\pi$} & \colhead{$\rm M_G$} & \colhead{$\rm G_{BP-RP}$} & \colhead{\Teff} & \colhead{log $g$} & \colhead{[Fe/H]} & \colhead{$\xi_{\rm t}$} &  \colhead{$R_V$} & \colhead{$v\sin i$} &
\colhead{A(Li)}\\
\colhead{} & \colhead{} & \multicolumn{2}{c}{(deg)}  & 
\multicolumn{2}{c}{(mas/yr)} & 
\colhead{(mas)} & 
\multicolumn{2}{c}{(mag)} & \colhead{(K)} & 
\colhead{} & 
\colhead{(dex)} &
\multicolumn{3}{c}{(km~s$^{-1}$)} & 
\colhead{(dex)}}

\startdata
WB105a & 170910006101001 & 71.32 & -65.31 & -17.39$\pm$0.01 & 61.57$\pm$0.02 & 3.92$\pm$0.01 & 3.93 & 0.79 & 5741$\pm$75 & 4.1$\pm$0.2 & -0.21$\pm$0.05 & 1.1 & 18.6$\pm$0.3 & 4.9$\pm$2.1 & 1.70 \\
WB105b & 161013004901159 & 71.31 & -65.32 & -17.49$\pm$0.01 & 61.57$\pm$0.02 & 3.96$\pm$0.01 & 6.52 & 1.25 & 4642$\pm$122 & 4.6$\pm$0.2 & -0.18$\pm$0.12 & 1 & 19.6$\pm$0.6 & 6.9$\pm$2.5 & 0.63 \\
WB106a & 150531000901057 & 206.89 & -78.93 & -6.24$\pm$0.01 & -10.12$\pm$0.01 & 1.85$\pm$0.01 & 4.42 & 0.85 & 6026$\pm$117 & 4.3$\pm$0.2 & -0.04$\pm$0.13 & 1.3 & -12.0$\pm$0.8 & 10.2$\pm$2.6 & 2.50 \\
WB106b & 150531000901043 & 206.91 & -78.92 & -6.29$\pm$0.01 & -10.07$\pm$0.02 & 1.83$\pm$0.01 & 5.00 & 0.96 & 5732$\pm$162 & 4.4$\pm$0.2 & -0.02$\pm$0.16 & 1.1 & -12.1$\pm$0.9 & 8.7$\pm$2.9 & 2.10 \\
WB107a & 170109002801101 & 111.23 & -59.66 & -3.45$\pm$0.02 & 10.21$\pm$0.02 & 4.12$\pm$0.01 & 4.65 & 0.81 & 5834$\pm$78 & 4.4$\pm$0.2 & 0.08$\pm$0.05 & 1.2 & 26.3$\pm$0.4 & 6.7$\pm$2.2 & 2.79 \\
WB107b & 150108002201269 & 111.23 & -59.67 & -3.29$\pm$0.01 & 9.73$\pm$0.01 & 4.13$\pm$0.01 & 5.76 & 1.05 & 5158$\pm$108 & 4.6$\pm$0.2 & 0.08$\pm$0.08 & 1 & 26.3$\pm$0.5 & 7.9$\pm$2.3 & 1.89 \\
WB108a & 160813003101358 & 310.22 & 1.55 & 29.37$\pm$0.01 & 3.88$\pm$0.01 & 2.96$\pm$0.01 & 4.55 & 0.77 & 5929$\pm$92 & 4.4$\pm$0.2 & -0.30$\pm$0.09 & 1.2 & -18.1$\pm$0.5 & 7.6$\pm$2.3 & 1.96 \\
WB108b & 160813003101356 & 310.22 & 1.58 & 29.44$\pm$0.01 & 3.81$\pm$0.01 & 2.97$\pm$0.01 & 5.00 & 0.83 & 5808$\pm$107 & 4.5$\pm$0.2 & -0.28$\pm$0.11 & 1.2 & -17.8$\pm$0.6 & 6.6$\pm$2.5 & 1.93 \\
\enddata
\tablecomments{The GALAH DR3 identifiers of each star are given in column 2. The coordinates are given in the column 3 and 4. The column 5 to 6 represent the proper motion at RA and DEC. The effective temperatures ($T_{\mathrm{eff}}$), the logarithmic of the surface gravity (log $g$) and the metallicity ([Fe/H]) with their errors in column 10, 11 and 12, respectively. The Microturbulence, the radial velocity, and the project rotational velocity are in the column 13 to 15. The A(Li) we measured are in the last column in this table. \textbf{The complete version of this table is available in the online Journal in the machine-readable table format.}}
\end{deluxetable*}